\documentclass[12pt]{article}
\usepackage{pgf}
\usepackage{tikz}
\usetikzlibrary{arrows,automata}
\usepackage{graphicx}
\usepackage{dcolumn}
\usepackage{bm}
\usepackage{amssymb}
\usepackage{amsmath}
\newcommand{\bq}{\begin{equation}}
\newcommand{\eq}{\end{equation}}

\newcommand{\bqr}{\begin{eqnarray}}
\newcommand{\eqr}{\end{eqnarray}}

\newcommand{\bqrx}{\begin{eqnarray*}}
\newcommand{\eqrx}{\end{eqnarray*}}

\newcommand{\br}{\begin{array}}
\newcommand{\er}{\end{array}}

\begin{document}

\pagestyle{empty}

\setlength{\parindent}{18pt}
\setlength{\footskip}{.5in}



\vspace*{.6in}
\begin{center}
From open quantum walks to unitary quantum walks
\end{center}
\begin{center}
Chaobin\, Liu  \footnote{cliu@bowiestate.edu, Department of Mathematics, Bowie State University\\Bowie, MD 20715 USA}

\end{center}
\begin{abstract}
We present an idea to convert to a unitary quantum walk any open quantum walk which is defined on lattices as well as on finite graphs. This approach generalizes to the domain of open quantum walks (or quantum Markov chains) the framework introduced by Szegedy for quantizing Markov chains. For the unitary quantum walks formulated in this article, we define the probability and the mean probability of finding the walk at a node, then derive the asymptotic mean probability.
 
\end{abstract}

keywords: Quantum Markov chains, open quantum walks, quantum walks.

\section{Introduction}


Markov chains or random walks (Markov chains on lattices or graphs), as statistical models of real-world processes, have broad applications in various fields of mathematics, computer science, physics and the natural sciences. An exact quantum extension of Markov chains (named {\it quantum Markov chains}) was recently presented by Gudder \cite{G2008}, at which the author defines a transition operation matrix as a matrix whose entries are completely positive maps whose column sums form a quantum operation. More recently, Attal et al. \cite{APS12, APSS12} introduced a formalism for discrete time {\it open quantum walks} (OQW), which are formulated as quantum Markov chains on lattices or graphs. For a short introduction to OQW and some of the recent developments on OQW (quantum Markov chains),  readers may refer to \cite{SF13}.

It is noteworthy that the concept of {\it quantum walks}, as a quantum counterpart of random walks,  has been proposed and well developed prior to the introduction of OQW. For a lively and informative elaboration of the history of such quantum walks and their connection to modern sciences, the reader is referred to \cite{ABNVW01, W2001, AAKV01, FG98, K03, CCDFGS03, Konno05, K06, AA2007, AMC2009, VA12} and references cited therein. 



Random walks (or Markov chains), in particular, can be used in computer science to formulate search algorithm. To invent a more efficient search algorithm, Szegedy \cite{S04} developed a generic method for quantizing classical algorithms based on random walks.  He shows that under certain conditions, the quantum version gives rise to a quadratic speed-up. Enlightened by the techniques (namely, quantizing a random walk to generate a unitary quantum walk) employed by Szegedy, it is possible, in the present context, to extend it to the domain of OQW. We will show how to convert open quantum walks to {\it unitary quantum walks}.

The structure of this article is outlined as follow: In section 2, we  briefly review the formalism of OQW, then we show how to convert an open quantum walk to a unitary quantum walk in section 3. Under the framework provided in section 3, we define the probability and the mean probability of the walk at nodes and discuss asymptotic distributions of the unitary quantum walks in section 4. In the following section, we show how to recover Szegedy's quantum walks by our approach. Finally we summarize and offer some remarks on the notions of Markov chains, open quantum walks and unitary quantum walks.





\section{Review of open quantum walks}

Before we proceed to show how to quantize an open quantum walk, let us to review briefly the formalism of open quantum walks introduced by Attal et al. \cite{APS12, APSS12}. One considers OQW on a directed graph $G(V, E)$. $V$ is the set of vertices of the graph $G$, $E$ is the set of oriented edge of $G$, $E=\{(j,k): j,k \in V\}$.  The position space of states corresponding to the dynamics on the graph is denoted by $\mathcal{H}_V=\mathbb{C}^{|V|}$ with the standard orthonormal basis $\{|j\rangle\}_{j\in V}$. The set of all linear operators on the Hilbert space $\mathcal{H}_V$ is denoted by $\mathfrak{B}(\mathcal{H}_V)$. The set of degrees of freedom (or called ``coin") is denoted by $C=\{c_1, c_2, ..., c_n\}$, the set of all linear operators on the Hilbert space $\mathcal{H}_C=\mathbb{C}^n=\mathrm{span}\{|c_1\rangle, |c_2\rangle, ..., |c_n\rangle \}$ is denoted by $\mathfrak{B}(\mathcal{H}_C)$.
\vskip 0.2in

Let $\mathfrak{D}(\mathcal{H}_C)\subset \mathfrak{B}(\mathcal{H}_C)$ denote the set of positive operators $\rho: \mathbb{C}^{n}\rightarrow \mathbb{C}^{n}$ with $\mathrm{Tr}(\rho)=1$. The operators $\rho$ are the so-called "density operators."  Thus, an overall state of the quantum walker can be described on the tensor product of the Hilbert spaces: $\mathfrak{B}(\mathcal{H}_C)\otimes \mathfrak{B}(\mathcal{H}_V).$

\vskip 0.1in
To describe the dynamics of the quantum walker, for each edge $(j,k)$, one introduces a bounded operator $B_j^k \in \mathfrak{B}(\mathcal{H}_C)$. This operator describe the change in the internal degree of freedom of the walker due to the shift from vertex $j$ to vertex $k$. By imposing for each vertex $j$ that 

\begin{equation}
\sum_k{B_j^k}^{\dagger}B_j^k=I_n,\label{relation_1}
\end{equation}

\noindent One makes sure, that for each vertex of the graph $j\in V$ there is a corresponding completely positive map (quantum operation) on the positive operators of $\mathfrak{B}(\mathcal{H}_C): \mathcal{M}_j(\tau)=\sum_k B_j^k\tau {B_j^k}^{\dagger}$. Since the operators $B_j^k$ act only on $\mathfrak{B}(\mathcal{H}_C)$ but don't perform transitions from vertex $j$ to vertex $k$, and operator $M_j^k\in \mathfrak{B}(\mathcal{H}_C)\otimes \mathfrak{B}(\mathcal{H}_V)$ is introduced with the form $M_j^k=B_j^k\otimes|k\rangle \langle j|$. It is evident that, with the Eq. (\ref{relation_1}) being satisfied, one has $\sum_{j,k}{M_j^k}^{\dagger}M_j^k=I$. This condition defines a completely positive map for a density operator $\rho\in \mathfrak{B}(\mathcal{H}_C)\otimes \mathfrak{B}(\mathcal{H}_V)$, i.e., 
$$\mathcal{M}(\rho)=\sum_k\sum_jM_j^k\rho M_j^k.$$

The above map defines the transition matrix for the open quantum walk. 

Given $\rho_0\in \mathfrak{B}(\mathcal{H}_C)\otimes \mathfrak{B}(\mathcal{H}_V)$, where $\mathrm{tr}(\rho_0)=1$, the expression $\rho_t=\mathcal{M}^t\rho_0$ is called the state of the walker at time $t$. The corresponding OQW with initial state $\rho_0$ is represented by the sequence $\{\rho_t\}_{t=0}^{\infty}$.

\section{How to convert an open quantum walk to a unitary quantum walk}

In what follows, we proceed to ``quantize" an open quantum walk. For any completely positive map $\mathcal{M}$ defined above, we will define a corresponding unitary quantum walk, residing on the augmented Hilbert space $\mathfrak{B}(\mathcal{H}_C)\otimes\mathcal{H}_V\otimes\mathcal{H}_V$ with the {\it Hilbert-Schmidt} inner product, defined by 
$$\langle A,B\rangle=\mathrm{tr}(A^{\dagger}B)$$. 

To define this walk, we first introduce the states:

\begin{equation}
|\psi_j\rangle:=\frac{1}{\sqrt{n}}\sum_k\sqrt{{B_j^k}^{\dagger}B_j^k}\otimes |j\rangle \otimes |k\rangle,\label{key_1}
\end{equation}

For $j=1, 2, ..., |V|$. Each such state is normalized, to see this, we calculate the norm of $|\psi_j\rangle$. $\||\psi_j\rangle\|=\sqrt{\langle\psi_j|\psi_j\rangle\rangle}=\sqrt{\frac{1}{n}\mathrm{tr}(\sum_k{B_j^k}^{\dagger}B_j^k)}=1$ by the condition given in Eq. (1). Then we define 
\begin{equation}
\Pi:=\sum_j^{|V|}|\psi_j\rangle \langle \psi_j| \label{key_pi}
\end{equation}

\noindent which is the projection on $\mathrm{span}\{|\psi_j\rangle:j=1, 2,..., |V|\}$, denoted by $\mathcal{H}_{\psi}$. This is a subspace of the augmented Hilbert space $\mathfrak{B}(\mathcal{H}_C)\otimes\mathcal{H}_V\otimes\mathcal{H}_V$. Finally let us define 

\begin{equation}
S:=\mathrm{I}_{n}\otimes\sum_{j,k=1}^{|V|}|j,k\rangle\langle k,j| \label{key_s}
\end{equation}

\noindent as the operator that swaps the two registers. 
\vskip 0.2in

With the operators defined above, a single step of the quantum walk is defined as the unitary operator $U:=S(2\Pi-1)$. Given $|\alpha_0\rangle \in \mathfrak{B}(\mathcal{H}_C)\otimes\mathcal{H}_V\otimes\mathcal{H}_V$, where 
$\||\alpha_0\rangle\|=1$, the expression $|\alpha_t\rangle=U^t|\alpha_0\rangle$ is called the state for the walk at time $t$. The corresponding quantum walk with initial state $|\alpha_0\rangle$ is represented by the sequence $\{|\alpha_t\rangle\}_{t=0}^{\infty}$.

\vskip 0.1in

\vskip 0.1in
In order to understand the behavior of a quantum walk, one needs to know the spectral properties of the unitary operator $U$. It will be helpful to begin with the study of an  $|V|\times |V|$ matrix $D=(d_{jk})$, as a linear transformation on the space $\mathcal{H}_V=\mathbb{C}^{|V|}$. The entries of this matrix is defined as follows:


\begin{equation}
d_{jk}=\frac{1}{n}\mathrm{tr}\bigg(\sqrt{{B_j^k}^{\dagger}B_j^k}\sqrt{ {B_k^j}^{\dagger}B_k^j}\bigg) \label{key_d}
\end{equation}

\vskip 0.1in

Let us then define an operator $A$ from the space $\mathcal{H}_V$ to $\mathcal{H}_{\psi}$:

\begin{equation}
A=\sum_{j=1}^{|V|}|\psi_j\rangle\langle j| \label{key_a}
\end{equation}

The following identities describe the relationships among these operators:

$$A^{\dagger}A=\mathbb{I}, AA^{\dagger}=\Pi, A^{\dagger}SA=D$$

Since $D$ is symmetric by its definition, without loss of the generality, we may assume that, via the Spectral Decomposition, $D=\sum_r \lambda_r|w_r\rangle \langle w_r|+\sum_{s}|u_s\rangle \langle u_s|-\sum_t|v_t\rangle \langle v_t\rangle$ where $\lambda_r\in (-1,1)$, and $\{|w_r\rangle, |u_s\rangle, |v_t\rangle\} $ is an {\it orthonormal} basis for $\mathcal{H}_V$.

\vskip 0.1in
Due to that $U|\psi_j\rangle=S|\psi_j\rangle$ and $US|\psi_j\rangle=2\sum_{k}d_{jk}S|\psi_k\rangle-|\psi_j\rangle$, we see that the subspace $\mathcal{H}_{\psi, S}=\mathrm{span}\{|\psi_j\rangle, S|\psi_j\rangle:j\in V\}$ is invariant under $U$. Notice that $\sum_r A|w_r\rangle\langle w_r| A^{\dagger}+\sum_s A|u_s\rangle\langle u_s| A^{\dagger}+\sum_t A|v_t\rangle\langle v_t|A^{\dagger}=AA^{\dagger}=\Pi$, the subspaces $\mathrm{span}\{A|w_r\rangle, A|u_s\rangle, A|v_t\rangle\}=\mathrm{span}\{|\psi_j\rangle:j\in V\}$ and thus $\mathcal{H}_{\psi, S}=\mathrm{span}\{A|w_r\rangle, SA|w_r\rangle, A|u_s\rangle, SA|u_s\rangle, A|v_t\rangle, SA|v_t\rangle\}$


\vskip 0.1in
It can be shown with not much difficulty that the following assertions hold:
\begin{enumerate}
\item  $A|w\rangle-e^{\pm i \arccos \lambda}SA|w\rangle$ is an eigenvector of $U$ with corresponding eigenvalue $e^{\pm i \arccos \lambda}$.
\item  $A|u\rangle=SA|u\rangle$, and $A|u\rangle$ is an eigenvector of $U$ with corresponding eigenvalue $1$.
\item $A|v\rangle=-SA|v\rangle$, and  $A|v\rangle $ is an eigenvector of $U$ with corresponding eigenvalue $-1$.
\end{enumerate}

\vskip 0.1in
The last two facts imply that $\mathcal{H}_{\psi, S}=\mathrm{span}\{A|w_r\rangle, SA|w_r\rangle, A|u_s\rangle, A|v_t\rangle\}$.

\vskip 0.1in
Since $\{A|w_r\rangle-e^{\pm i \arccos \lambda_r}SA|w_r\rangle, A|u_s\rangle, A|v_t\rangle\}$ forms an orthogonal set, $\mathcal{H}_{\psi, S}=\mathrm{span}\{A|w_r\rangle-e^{\pm i \arccos \lambda_r}SA|w_r\rangle, A|u_s\rangle, A|v_t\rangle\}$. After normalizing each vector in this orthogonal set, we obtain an orthonormal basis for the invariant subspace $\mathcal{H}_{\psi, S}$, which is given by the set $\{A|w_r^+\rangle, A|w_r^-\rangle, A|u_s\rangle, A|v_t\rangle\}$. Here 

\begin{enumerate}
  \item  $A|w_r^+\rangle=(A|w_r\rangle-e^{i \arccos \lambda_r}SA|w_r\rangle/\sqrt{2-2\lambda_r^2}$
  \item  $A|w_r^-\rangle=(A|w_r\rangle-e^{-i \arccos \lambda_r}SA|w_r\rangle)/\sqrt{2-2\lambda_r^2}$
\end{enumerate}
\vskip 0.1in
To summarize the arguments made above, $\mathcal{H}_{\psi, S}=\mathrm{span}\{|\psi_j\rangle, S|\psi_j\rangle:j\in V\}$ as an invariant subspace under $U$, can be recasted by 

$$\mathrm{span}\{A|w_r^+\rangle, A|w_r^-\rangle, A|u_s\rangle, A|v_t\rangle\}$$

\noindent where this spanning set is the collection of the orthonormal eigenvectors of $U$ associated with the key operator $D$.

\vskip 0.1in
Let us decompose the Hilbert space $\mathfrak{B}(\mathcal{H}_C)\otimes\mathcal{H}_V\otimes\mathcal{H}_V$ into $\mathcal{H}_{\psi,S}$ and its orthogonal complement $\mathcal{H}_{\psi,S}^{\bot}$, i.e., $\mathfrak{B}(\mathcal{H}_C)\otimes\mathcal{H}_V\otimes\mathcal{H}_V=\mathcal{H}_{\psi,S}\oplus\mathcal{H}_{\psi, S}^{\bot}$. It is easy to check that the actions of $U$ and $U^2$ on $\mathcal{H}_{\psi,S}^{\bot}$ are $-S$ (thus $\mathcal{H}_{\psi, S}^{\bot}$ is invariant under $U$) and the identity $\mathbb{I}$, respectively. Therefore, the nontrivial action of $U$ only takes place on the subspace $\mathcal{H}_{\psi,S}$ of a dimension less than or equal to $2|V|$ (this maximum dimension can be achieved only if $D$ does not have both $1$ and $-1$ as its eigenvalues). Based on the aforesaid observation, we may confine the initial state of the quantum walk to the subspace $\mathcal{H}_{\psi, S}$, which is spanned by the set of the orthonormal eigenvectors of $U$: $\{A|w_r^+\rangle, A|w_r^-\rangle, A|u_s\rangle, A|v_t\rangle\}$.

\section{Asymptotic distribution of the quantum walks}

We now turn to study the evolution of the quantum walk. Beginning with the initial state $|\alpha_0\rangle$, the state of the unitary quantum walk at time $t$ is $|\alpha_t\rangle=U^{t}|\alpha_0\rangle$. Since $U$ is unitary, in general the limit $\lim_{t\rightarrow \infty} |\alpha_t\rangle$ does not exist. Now consider instead the probability distribution on the states of the underlying open quantum walks induced by $|\alpha_t\rangle$,

Definition. $P_t(j|\alpha_0)=\sum_{k}\langle \mathbb{I}_{n^2}\otimes |j,k\rangle\langle j,k|\alpha_t\rangle\langle\alpha_t|\rangle$. Here $P_t(j|\alpha_0)$ is the probability of finding the walk at the node $v_j$ at time $t$.

As a matter of fact, $P_t$ usually do not converge either. However, the average of $P_t$ over time is convergent. We define:

Definition. $\overline{P_T}(j|\alpha_0)=\frac{1}{T}\sum_{t=1}^T P_t(j|\alpha_0)$. This is the mean probability of finding the walk at the node $v_j$ over time interval $[1,T]$.

For the sake of brevity, we denote the set of eigenvalues of $U$ by $\{\phi_l\}$, and the set of the corresponding eigenvalues of $U$ by $\{\mu_l\}$. Via a routine reasoning, we can arrive at a theorem regarding the asymptotic distribution of the unitary quantum walks.


Theorem\, Given an open quantum walk on the state space $V$ with the transition matrix $\mathcal{M}$, the induced quantum walk is defined as $|\alpha_t\rangle=U^{t}|\alpha_0\rangle$ where the initial state $|\alpha_0\rangle=\sum_l\langle \phi_l|\alpha_0\rangle |\phi_l\rangle$, then 
$$\lim_{T\mapsto \infty}\overline{P_T}(j|\alpha_0)=\sum_k\sum_{l, m}\langle \phi_l|\alpha_0\rangle \langle \alpha_0| \phi_m\rangle \langle \mathbb{I}_n\otimes |j,k\rangle\langle j,k|,|\phi_l\rangle\langle\phi_m|\rangle$$
where the first sum is over all values of $k$, and the second sum is only on pairs $l, m$ such that $\mu_l=\mu_m$.

\vskip 0.2in

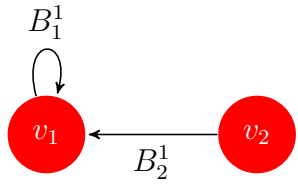
\begin{figure}
\begin{center}

\begin{tikzpicture}[->,>=stealth',shorten >=1pt,auto,node distance=2.8cm,
                    semithick]
  \tikzstyle{every state}=[fill=red,draw=none,text=white]

  \node[state] (A)                    {$v_2$};
  \node[state]         (B) [left of=A] {$v_1$};

  \path (A) edge              node {$B_2^1$} (B)
        (B) edge [loop above] node {$B_1^1$} (B);
        
\end{tikzpicture}
\end{center}
\caption{Open quantum walk}\label{fig:oqw}
\end{figure}

\noindent {\bf Example}\,\, Let us consider a simple open quantum walk on the graph with two vertices (see Figure \ref{fig:oqw}),  its transition operator is given by

\begin{equation}
M= \left[\begin{array}{cc}
B_1^1& B_2^1\\
0     & 0
\end{array}\right]\,\label{eqnU_{ec}}
\end{equation}

where \begin{equation}
B_1^1=\left[\begin{array}{cc}
0& 1\\
1    & 0
\end{array}\right],\,\,\,\label{}
B_2^1=\left[\begin{array}{cc}
1& 0\\
0    & -1
\end{array}\right]\,.\label{}
\end{equation}
Case 1. when \begin{equation}
\rho_0=\left[\begin{array}{cc}
\frac{1}{4}& 0\\
0    & \frac{1}{4}
\end{array}\right]\otimes |1\rangle\langle 1|+
\left[\begin{array}{cc}
\frac{1}{4}& 0\\
0    & \frac{1}{4}
\end{array}\right]\otimes|2\rangle\langle 2|\,,\label{}
\end{equation}

then 
\begin{equation}
\rho_{\infty}=\left[\begin{array}{cc}
\frac{1}{2}& 0\\
0    & \frac{1}{2}
\end{array}\right]\otimes|1\rangle\langle1|
\end{equation}

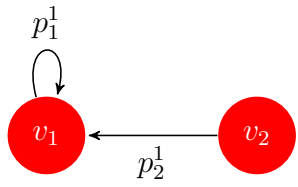
\begin{figure}
\begin{center}

\begin{tikzpicture}[->,>=stealth',shorten >=1pt,auto,node distance=2.8cm,
                    semithick]
  \tikzstyle{every state}=[fill=red,draw=none,text=white]

  \node[state] (A)                    {$v_2$};
  \node[state]         (B) [left of=A] {$v_1$};

  \path (A) edge              node {$p_2^1$} (B)
        (B) edge [loop above] node {$p_1^1$} (B);
        
\end{tikzpicture}
\end{center}
\caption{Markov chain}  \label{fig:MC}
\end{figure}

Case 2. When \begin{equation}
\rho_0=\left[\begin{array}{cc}
\frac{3}{4}& 0\\
0    & \frac{1}{4}
\end{array}\right]\otimes |2\rangle\langle 2|\,, \label{}
\end{equation}

then 
\begin{equation}
\rho_{2k-1}=\left[\begin{array}{cc}
\frac{3}{4}& 0\\
0    & \frac{1}{4}
\end{array}\right]\otimes|1\rangle\langle1|,\, \mathrm{or}\,\,
\rho_{2k}=\left[\begin{array}{cc}
\frac{1}{4}& 0\\
0    & \frac{3}{4}
\end{array}\right]\otimes|1\rangle\langle 1|\,.\label{}
\end{equation}
Here $k=1, 2, ...$. Therefore $\rho_{\infty}$ does not exist in this case! However, if one performs measurements of the position of the walker at each node, the quantum trajectories \cite{APSS12} of the walker will converge to $p_{\infty}(v_{1})=1, p_{\infty}(v_2)=0$ in each case. These quantum trajectories {\it imitate} the corresponding classical Markov chain (illustrated in Figure \ref{fig:MC}) with transition matrix given by

\begin{equation}
P= \left[\begin{array}{cc}
1& 1\\
0     & 0
\end{array}\right]\,\label{transition P}
\end{equation}

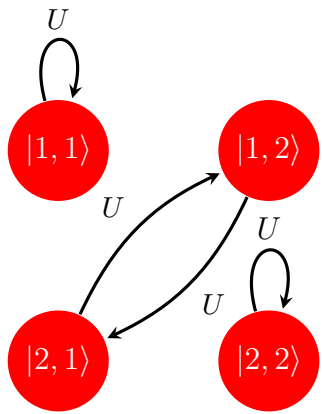
\begin{figure}
\begin{center}
\begin{tikzpicture}[->,>=stealth,shorten >=1pt,auto,node distance=2.8cm, semithick] 
\tikzstyle{every state}=[fill=red,draw=none,text=white] 
 
\node[state]      (1) {$|1,1\rangle$};
\node[state]         (2) [right of=1] {$|1,2\rangle$};
\node[state]         (3) [below of=1] {$|2,1\rangle$}; 
\node[state]         (4) [below of=2] {$|2,2\rangle$}; 
 
\path 
(1) edge[line width=1.20pt,loop above] node {\small $U$} (1) 
(2) edge[line width=1.20pt,bend left=20] node {\small $U$} (3) 
(3) edge[line width=1.20pt,bend left=20] node {\small $U$} (2) 
(4) edge[line width=1.20pt,loop above] node {$U$} (4) 
 ;

\end{tikzpicture} 
\caption{Quantum walk} \label{Quantum walk}
\end{center}
\end{figure}

\vskip 0.2in
Under the framework described in section 3, we will convert this OQW to a unitary quantum walk (see Figure \ref{Quantum walk}).

By Eq.(\ref{key_1}), 

\begin{equation}
|\psi_1\rangle=\frac{1}{\sqrt{2}}\left[\begin{array}{cc}
1& 0\\
0    & 1
\end{array}\right]\otimes |1,1\rangle,\,\,\,\label{}
|\psi_2\rangle=\frac{1}{\sqrt{2}}\left[\begin{array}{cc}
1& 0\\
0    & 1
\end{array}\right]\otimes |2,1\rangle,\,\,\,\label{}
\end{equation}

Case 1. We choose $|\alpha_0\rangle=\frac{1}{\sqrt{2}}(|\psi_1\rangle+|\psi_2\rangle)=\frac{1}{2}\mathbb{I}_2\otimes|1,1\rangle+\frac{1}{2}\mathbb{I}_2\otimes |2,1\rangle$, then we have 

$|\alpha_1\rangle=U|\alpha_0\rangle=\frac{1}{2}\mathbb{I}_2\otimes|1,1\rangle+\frac{1}{2}\mathbb{I}_2\otimes |1,2\rangle$, and $P_1(1|\alpha_0)=1, P_1(2|\alpha_0)=0$.

$|\alpha_2\rangle=U|\alpha_1\rangle=\frac{1}{2}\mathbb{I}_2\otimes|1,1\rangle-\frac{1}{2}\mathbb{I}_2\otimes |2,1\rangle$,  and $P_2(1|\alpha_0)=P_2(2|\alpha_0)=\frac{1}{2}$.

$|\alpha_3\rangle=U|\alpha_2\rangle=\frac{1}{2}\mathbb{I}_2\otimes|1,1\rangle-\frac{1}{2}\mathbb{I}_2\otimes |1,2\rangle$, and $P_3(1|\alpha_0)=1, P_3(2|\alpha_0)=0$.

$|\alpha_4\rangle=U|\alpha_3\rangle=\frac{1}{2}\mathbb{I}_2\otimes|1,1\rangle+\frac{1}{2}\mathbb{I}_2\otimes |2,1\rangle$,  and $P_4(1|\alpha_0)=P_4(2|\alpha_0)=\frac{1}{2}$.

\vskip 0.2in
It is seen that the states of the quantum walker $\{\alpha_t\}_{t=0}^{\infty}$ are periodic, and the period is 4. The limiting distribution is: $\overline{P}_{\infty}(1|\alpha_0)=\frac{3}{4}$ and $\overline{P}_{\infty}(2|\alpha_0)=\frac{1}{4}$.

Case 2. We choose $|\alpha_0\rangle=\frac{\sqrt{2}}{2}\mathbb{I}_2\otimes |1,2\rangle$, then we have 

$|\alpha_1\rangle=U|\alpha_0\rangle=-\frac{\sqrt{2}}{2}\mathbb{I}_2\otimes |2,1\rangle$, and $P_1(1|\alpha_0)=0, P_1(2|\alpha_0)=1$.

$|\alpha_2\rangle=U|\alpha_1\rangle=-\frac{\sqrt{2}}{2}\mathbb{I}_2\otimes|1,2\rangle$,  and $P_2(1|\alpha_0)=1, P_2(2|\alpha_1)=0$.

$|\alpha_3\rangle=U|\alpha_2\rangle=\frac{\sqrt{2}}{2}\mathbb{I}_2\otimes|2,1\rangle$, and $P_3(1|\alpha_0)=0, P_3(2|\alpha_0)=1$.

$|\alpha_4\rangle=U|\alpha_3\rangle=\frac{\sqrt{2}}{2}\mathbb{I}_2\otimes|1,2\rangle$,  and $P_4(1|\alpha_0)=1, P_4(2|\alpha_0)=0$.

\vskip 0.2in
Again the the states of the quantum walker are periodic, and the period is also equal to 4. However, the limiting distribution is different from the one in case 1. In this case, it is: $\overline{P}_{\infty}(1|\alpha_0)=\overline{P}_{\infty}(2|\alpha_0)=\frac{1}{2}$.

In summary, the mean probability distributions of the quantum walks are convergent by Theorem 1. The asymptotic mean distributions are dependent on initial states because the underlying network is not connected in the example we discussed. In general, it is speculated that the asymptotic distribution for the quantum walk may be unique if the underlying network is strongly connected.

\section{Recovering Szegedy's quantum walks}

Let us recall how Szegedy \cite{S04} quantizes a random walk (Markov chain) to produce a unitary quantum walk (we call it Szegedy's quantum walk).

\vskip 0.1in

A discrete-time random walk on an $N$-vertex graph can be represented by an $N\times N$ matrix
$P$ in which the entry $p_{kj}$ represents the probability of making a transition to $k$ from $j$. Let $u$ be the probability vector which represents the starting distribution. Then the probability distribution after one step of the walk becomes $Pu$. To preserve normalization,
we must have $\sum_{k=1}^N p_{kj} = 1$, such a matrix is often said to be stochastic.

\vskip 0.1in

For the $N\times N$ stochastic matrix $P$, one can define a corresponding
discrete-time quantum walk by a unitary operation on the Hilbert space $\mathbb{C}^N\otimes \mathbb{C}^N$. To define this unitary operation, one first introduces the states

$$|\psi_j\rangle:=\sum_{k=1}^N\sqrt{p_{kj}}|j,k\rangle$$
\noindent for $j=1, 2, ..., N$. Each such state is normalized because $P$ is stochastic. Then one defines 

$$\Pi:=\sum_{j=1}^N |\psi_j\rangle\langle\psi_j|$$
\noindent which is the projection onto $\mathrm{span}\{\psi_j\rangle:j=1, 2, ..., N\}$, and finally one lets 

$$S:=\sum_{j,k=1}^N |j,k\rangle\langle k,j|$$

\noindent be the operator that swaps the two registers. Then a single step of Szegedy's quantum walk is defined as
the unitary operator $U:=S(2\Pi-1)$. 

To see how the operator $U$ acts on the underlying Hilbert space $\mathcal{H}_V\otimes\mathcal{H}_V=\mathrm{span}\{|j,k\rangle: j,k \in V \}$, it suffices to check $U(|j_0,k_0\rangle)$. By the definition for $U$ defined above, we have

\begin{equation}
U(|j_0, k_0\rangle)=2\sqrt{p_{k_0j_0}}\sum_{k}\sqrt{p_{kj_0}}|k,j_0\rangle-|k_0,j_0\rangle \label{szegedy}
\end{equation}

\vskip 0.1in

Now we turn to the setup of converting an open quantum walks to a unitary quantum walks presented in this article. 

Consider the stochastic transition matrix $P=(p_{kj})$, and  any family of unitary
operators $U_j^k$ on $\mathbb{C}^n$ for $j,k\in V$, we set 

$$B_j^k=\sqrt{P_{kj}}U_j^k,$$

It is seen that $\sum_{k=1}^{|V|}{B_j^k}^{\dagger}B_j^k=\mathbb{I}_n$ for all $j$.

\vskip 0.2in
Employing the unitary operation $U$ defined in section 3, it can be calculated that 

\begin{equation}
U(\mathbb{I}_n\otimes |j_0, k_0\rangle)=\mathbb{I}_n\otimes[2\sqrt{p_{k_0j_0}}\sum_{k}\sqrt{p_{kj_0}}|k,j_0\rangle-|k_0,j_0\rangle] \label{I}
\end{equation}

From Eqs. (\ref{szegedy}) and (\ref{I}), we can see that the two unitary operators are identical over $\mathcal{H}_V\otimes\mathcal{H}_V$. It is noted that the normed spaces $\mathcal{H}_V\otimes\mathcal{H}_V$ (the underlying space for Szegedy's walk) and $\mathrm{span}\{\mathbb{I}_n\}\otimes \mathcal{H}_V\otimes\mathcal{H}_V$ (a subspace of the augmented Hilbert space $\mathfrak{B}(\mathcal{H}_C)\otimes\mathcal{H}_V\otimes\mathcal{H}_V)$ are isometrically isomorphic, therefore Szegedy's quantum walk can be considered as a special case of the  quantum walk defined in this article.



\section{Summery and some remarks}

We present an scheme to convert an open quantum walk to a unitary quantum walk. This approach extends to the domain of open quantum walks (or quantum Markov chains) the framework introduced by Szegedy \cite{S04} for quantizing Markov chains. Open quantum walks can be viewed as an exact quantum extension of random walks (Markov chains). To illustrate the relationships among the notions of random walks, quantum walks and open quantum walks, we draw a diagram below (the work done in this article is marked in grey).


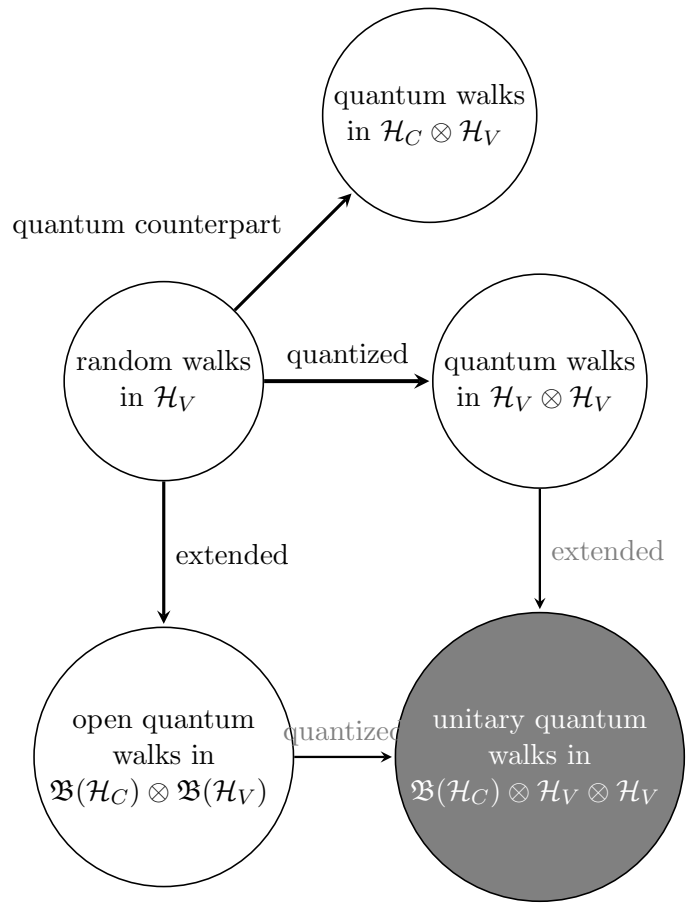
\begin{center}
\begin{tikzpicture}[->,>=stealth,shorten >=1pt,auto,node distance=5.0cm, semithick] 
\tikzstyle{every state}=[fill=gray,draw=black,text=white] 
 
\node[state, fill=white, text=black, inner sep = -3pt]      (1) {\small \begin{tabular}{c} 
    random walks \\ 
in $\mathcal{H}_V$
    {\footnotesize } 
\end{tabular}}; 
\node[state, fill=white,text = black, inner sep = -3pt]         (2) [right of=1] {\small \begin{tabular}{c} 
   quantum walks \\
in $\mathcal{H}_V\otimes\mathcal{H}_V$ 
    {\footnotesize } 
\end{tabular}}; 
\node[state, fill=white,text=black, inner sep = -3pt]         (3) [below of=1] {\small \begin{tabular}{c} 
    open quantum \\
walks in \\
$\mathfrak{B}(\mathcal{H}_C)\otimes\mathfrak{B}(\mathcal{H}_V)$
    {\footnotesize }
\end{tabular}}; 
\node[state, fill=gray, inner sep = -3pt]         (4) [below of=2] {\small \begin{tabular}{c} 
    unitary quantum \\
walks in \\
$\mathfrak{B}(\mathcal{H}_C)\otimes\mathcal{H}_V\otimes\mathcal{H}_V$ 
    {\footnotesize } 
\end{tabular}};
\node[state, fill=white, text=black, inner sep = -3pt]         (5) [above right of=1] {\small \begin{tabular}{c} 
    quantum walks \\
in $\mathcal{H}_C\otimes\mathcal{H}_V$ 
    {\footnotesize } 
\end{tabular}};
\path 
(1) edge[line width=1.5pt] node {\small quantized} (2) 
(1) edge[line width=1.20pt] node {\small extended} (3) 
(1) edge[line width=1.20pt] node {\small quantum counterpart} (5) 
(2) edge[line width=0.87pt] node {\small \color{gray}extended} (4) 
(3) edge[line width=0.87pt] node {\small \color{gray}quantized} (4) 
 ; 
\end{tikzpicture}
\end{center}

\vskip 0.2in
For the quantum walks we introduced in this article, we define the probability and the mean probability of finding the walker at a node, then we obtain a theorem regarding the asymptotic mean probability distribution.

\end{document}